\documentclass[aps,longbibliography,notitlepage,floatfix,11pt,tightenlines,nofootinbib]{revtex4-1}

\usepackage{amsfonts,amsmath,amssymb,bm}
\usepackage{graphicx,rotating,xcolor}
\usepackage{hyperref}
\usepackage[margin=1in]{geometry}

\providecommand{\norm}[1]{\left\| #1 \right\|}

\providecommand{\bmru}[1]{\bm{\mathrm{#1}}}

\begin{document}

\title{Exterior dissipation, proportional decay, and integrals of motion}
\author{M. Aureli}
	\email{maureli@unr.edu}
\author{J. A. Hanna}
	\email{jhanna@unr.edu}
\affiliation{Department of Mechanical Engineering, University of Nevada, Reno, NV 89557, U.S.A.}
	\date{\today}
	
\begin{abstract}
Given a dynamical system with $m$ independent conserved quantities, we construct a multi-parameter family of new systems in which these quantities evolve monotonically and proportionally, and are replaced by $m-1$ conserved linear combinations of themselves, with any of the original quantities as limiting cases.
The modification of the dynamics employs an exterior product of gradients of the original quantities, and often evolves the system towards asymptotic linear dependence of these gradients in a nontrivial state.
The process both generalizes and provides additional structure to existing techniques for selective dissipation in the literature on fluids and plasmas, nonequilibrium thermodynamics, and nonlinear controls.
It may be iterated or adapted to obtain any reduction in the degree of integrability.  It may enable % selective control or energy transfer, 
discovery of extremal states, limit cycles, or solitons, and the construction of new integrable systems from superintegrable systems.
We briefly illustrate the approach by its application to the cyclic three-body Toda lattice, driven from an aperiodic orbit towards a limit cycle. 
\end{abstract}

\maketitle

%%%%%%%%%%%%%%%%%%%%%%%%%%%%%%%%%%%%%%%%%%%%%%%%%%%%%%%%%%%%%%%%%%%%%%

A fundamental description of classical mechanics is of a conservative, although non-integrable, dynamical system.  Realistically, one can only account for the behavior of an open subsystem, whose interactions with the external world lead to nonconservative effects.
It is often of interest to observe the time evolution of such a system when only certain of its ideally conserved quantities are made to change, perhaps monotonically.
%symmetries broken
Modifications of dynamics, such as the ``metriplectic'' and ``double bracket'' methods, that selectively and monotonically violate a conservation law have been developed in efforts towards a unified framework for conservative and dissipative systems \cite{Kaufman84, Morrison84, Morrison86, Morrison09, Grmela86},
including nonequilibrium thermodynamics 
\cite{EdwardsBeris91, GrmelaOttinger97, Gay-BalmazYoshimura20}, 
from the desire to realistically model or search for extremal states of geophysical or magnetohydrodynamic flows 
\cite{Vallis89, CarnevaleVallis90, Shepherd90-2, Gay-BalmazHolm14, Morrison17}, 
and in control and optimization settings 
\cite{Brockett91, Bloch96}. 
In these examples, only one quantity, such as the energy or enstrophy, is modified, while all the others are held fixed.
While it is possible to combine multiple nonconservative terms of this type, no clear additional structure is found in the resulting trajectories and, presumably, if all quantities are dissipated, nothing is conserved and the system will be driven to a trivial (motionless) equilibrium state.
%decay

In this note, we show that processes leading to the limited outcome of constrained extremization of single quantities can instead be viewed as end points of a continuum of modified dynamical systems that preserve as much integrability as possible of the original dynamics.
%is in some sense maximally conservative
A geometrically motivated construction incorporating the exterior product of gradients conserves $m-1$ linear combinations of an original $m$ independent integrals that now evolve monotonically and proportionally, their rates given by a set of tunable coefficients and scaling with the hypervolume of an $m$-parallelotope spanned by the gradients.
When any of these gradients become linearly dependent, nontrivial states of the original dynamics are obtained.  
We also provide an equivalent, but very simple and numerically convenient, linear-algebraic expression for the modified dynamics.
%While writing up this note, it came to our attention that some aspects of our initial thought process and construction can be seen in the approach of Shashikanth \cite{Shashikanth10} who, however, like the above-cited works, considered only terms to dissipate a single quantity and thus did not discover the new collective phenomena and conservations we present here.
While writing, it came to our attention that some aspects of our initial thought process and construction can be seen in \cite{Shashikanth10}, although like other previous approaches, this was applied to dissipate only a single quantity and thus did not reveal the collective phenomena and conservations we present here.

Consider a dynamics $\dot{\bmru{x}} = \bmru{f}(\bmru{x})$ for $\bmru{x} \in \mathbb{R}^n$.  Let this system admit $m\le n$ independent first integrals, time-invariant constants of the motion $Q_i(\bmru{x})$, $i \in \{1,\ldots,m\}$.  Thus, the total time derivative of any $Q_i$ vanishes, $\dot Q_i(\bmru{x}) = d_{\bmru{x}} Q_i \cdot \bmru{f} =0$.
%(or Lie derivative, indicated below with the symbol $L_{\bmru{f}}[\bullet]$) is zero, i.e., by definition~\cite{Arnold1989Mathematical}
%\begin{equation}\label{eq:11}
%L_{\bmru{f}}[ Q_i(\bmru{x}) ]= \partial_{\bmru{x}} Q_i \cdot \bmru{f} =0
%\end{equation}
%\begin{equation}\label{eq:11}
%	\dot Q_i(\bmru{x}) = \partial_t Q_i + d_{\bmru{x}} Q_i \cdot \bmru{f} =0 \, .
%\end{equation}
Now consider a modification of the original dynamics, so that 
\begin{equation}
	\dot{\bmru{x}} = \bmru{f}(\bmru{x}) + \bmru{d} \, ,
\end{equation}
and the integrals change as
\begin{equation}\label{eq:qi_dot}
	\dot Q_i = d_{\bmru{x}} Q_i \cdot \bmru{d} \equiv \bmru{v}_i  \cdot \bmru{d} \ne 0 \, .
\end{equation}
%$\dot{\bmru{x}} = \tilde{\bmru{f}}(\bmru{x},t) = \bmru{f}(\bmru{x},t) + \bmru{d}$. The first integrals become
%\begin{equation}\label{eq:12}
%L_{\tilde{\bmru{f}}}[ Q_i(\bmru{x}) ] = \partial_{\bmru{x}} Q_i \cdot \bmru{d} \ne 0
%\end{equation}
For notational convenience we have denoted the $m$ gradients $\in \mathbb{R}^n$ of the $m$ independent quantities $Q_i$ as $\bmru{v}_i \equiv d_{\bmru{x}} Q_i$.  These gradients will be linearly independent.
 The vector $\bmru{d}$ may be envisioned as a dissipation due to contact with an external natural system, or as a control input that seeks to amplify or decay $\bmru{x}$ or some $Q_i$.

It is of interest to construct a $\bmru{d}$ that will selectively amplify or dissipate one of the $Q_i$, so that this quantity changes monotonically while all the others are unaffected. 
Accordingly, we seek a generalization of the double cross products of two vectors employed in \cite{Hannaintegrable}. 
 To effect this construction, we first recall the exterior (wedge) product \cite{Bishop1980Tensor} of the $m$ vectors $\bmru{v}_i$ as the complete antisymmetrization of their tensor product (here denoted in dyadic notation by juxtaposition), % $\otimes$,
\begin{equation}\label{eq:wedge}
  \bmru{v}_1 \wedge \bmru{v}_2 \wedge \ldots \wedge \bmru{v}_m = \frac{1}{m!}\sum _{\sigma} \mathrm{sgn}(\sigma)\, \bmru{v}_{\sigma(1)} \bmru{v}_{\sigma(2)} \ldots \bmru{v}_{\sigma(m)} \, ,
%  \bmru{v}_1 \wedge \bmru{v}_2 \wedge \ldots \wedge \bmru{v}_m = \frac{1}{m!}\sum _{\sigma} \mathrm{sgn}(\sigma)\, \bmru{v}_{\sigma(1)} \otimes \bmru{v}_{\sigma(2)} \otimes \ldots \otimes \bmru{v}_{\sigma(m)} \, ,
\end{equation}
with the sum taken over all possible permutations $\sigma$ of $\{1,2,\ldots,m\}$, each with parity $\mathrm{sgn}(\sigma)$.  %The symbol $\otimes$ indicates the usual tensor product.
We next introduce a contraction operation with a tensor product to form a bracket defined as
\begin{equation}\label{eq:bracket}
  [[Q_1,Q_2,\ldots,Q_m]]_i =
      (\bmru{v}_i \wedge \bmru{v}_1 \wedge \ldots \wedge \bmru{v}_{i-1} \wedge \bmru{v}_{i+1} \wedge \ldots \wedge \bmru{v}_m ) \overset{m-1}{\cdot} (\bmru{v}_1 \ldots \bmru{v}_{i-1} \bmru{v}_{i+1} \ldots \bmru{v}_m) \, ,
%  [[Q_1,Q_2,\ldots,Q_m]]_i =
%      (\bmru{v}_i \wedge \bmru{v}_1 \wedge \ldots \wedge \bmru{v}_{i-1} \wedge \bmru{v}_{i+1} \wedge \ldots \wedge \bmru{v}_m ) \overset{m-1}{\cdot} (\bmru{v}_1\otimes \ldots \otimes \bmru{v}_{i-1} \otimes \bmru{v}_{i+1} \otimes \ldots \otimes \bmru{v}_m) \, ,
\end{equation}
%\begin{equation}\label{eq:bracket}
%  [[\bmru{v}_1,\bmru{v}_2,\ldots,\bmru{v}_m]]_i =
%      (\bmru{v}_i \wedge \bmru{v}_1 \wedge \ldots \wedge \bmru{v}_{i-1} \wedge \bmru{v}_{i+1} \wedge \ldots \wedge \bmru{v}_m ) \overset{m-1}{\cdot} (\bmru{v}_1\otimes \ldots \otimes \bmru{v}_{i-1} \otimes \bmru{v}_{i+1} \otimes \ldots \otimes \bmru{v}_m) \, ,
%\end{equation}
where the subscripted index $i$ on the bracket indicates one special vector from the collection of gradients, to be placed at the left of the exterior product that will form a fully antisymmetric $m$-tensor, and removed from the $(m-1)$-tensor product with which this is contracted to form a vector.  The contraction between the $(m-1)$ vectors on the right of the exterior product and the entire tensor product is to be performed in parallel, rather than reflected, order. %sequentially (rather than in reflected order). %Here, the convention for the repeated dot product is such that $(\bmt{A} \overset{m-1}{\cdot} \bmt{V})_{i_1}  = (\bmt{A})_{i_1i_2\ldots i_m}   (\bmt{V})_{ i_2\ldots i_m}$, with implied summation over the repeated indices. 
%This means that, for example, $[\bmru{v}_1,\bmru{v}_2]^1 = (\bmru{v}_1 \wedge \bmru{v}_2)\cdot \bmru{v}_2 = \frac{1}{2}( \norm{\bmru{v}_2}^2 \bmru{v}_1 - (\bmru{v}_1\cdot \bmru{v}_2)\bmru{v}_2)$ 
For example, 
%$[[\bmru{v}_1,\bmru{v}_2,\bmru{v}_3]]_2$
$[[Q_1,Q_2,Q_3]]_2 = (\bmru{v}_2 \wedge \bmru{v}_1 \wedge\bmru{v}_3) : ( \bmru{v}_1 \bmru{v}_3) 
= \frac{1}{6} \Big( \Big[ \norm{\bmru{v}_1}^2 \norm{\bmru{v}_3}^2 - (\bmru{v}_3 \cdot \bmru{v}_1)^2 \Big] \bmru{v}_2\\
 + \Big[-(\bmru{v}_2\cdot\bmru{v}_1)\norm{\bmru{v}_3}^2 
 + (\bmru{v}_3 \cdot \bmru{v}_1) (\bmru{v}_2 \cdot \bmru{v}_3 ) \Big] \bmru{v}_1
 + \Big[ (\bmru{v}_2 \cdot \bmru{v}_1) (\bmru{v}_1 \cdot \bmru{v}_3)   
  - \norm{\bmru{v}_1}^2(\bmru{v}_2\cdot\bmru{v}_3) \Big] \bmru{v}_3 \Big)$.  
%$[[Q_1,Q_2,Q_3]]_2 = (\bmru{v}_2 \wedge \bmru{v}_1 \wedge\bmru{v}_3) : ( \bmru{v}_1 \otimes\bmru{v}_3) 
%= \frac{1}{6} \Big( \Big[ \norm{\bmru{v}_1}^2 \norm{\bmru{v}_3}^2 - (\bmru{v}_3 \cdot \bmru{v}_1)^2 \Big] \bmru{v}_2
% + \Big[-(\bmru{v}_2\cdot\bmru{v}_1)\norm{\bmru{v}_3}^2 
% + (\bmru{v}_3 \cdot \bmru{v}_1) (\bmru{v}_2 \cdot \bmru{v}_3 ) \Big] \bmru{v}_1
% + \Big[ (\bmru{v}_2 \cdot \bmru{v}_1) (\bmru{v}_1 \cdot \bmru{v}_3)   
%  - \norm{\bmru{v}_1}^2(\bmru{v}_2\cdot\bmru{v}_3) \Big] \bmru{v}_3 \Big)$.  
%\begin{align}
%	[[\bmru{v}_1,\bmru{v}_2,\bmru{v}_3]]_2 &= (\bmru{v}_2 \wedge \bmru{v}_1 \wedge\bmru{v}_3) : ( \bmru{v}_1 \otimes\bmru{v}_3) \nonumber \\ 
%&= \frac{1}{6}\left( \left[ \norm{\bmru{v}_1}^2 \norm{\bmru{v}_3}^2 - (\bmru{v}_3 \cdot \bmru{v}_1)^2 \right] \bmru{v}_2 \right. \nonumber \\
%&\quad + \left[-(\bmru{v}_2\cdot\bmru{v}_1)\norm{\bmru{v}_3}^2 
% + (\bmru{v}_3 \cdot \bmru{v}_1) (\bmru{v}_2 \cdot \bmru{v}_3 ) \right] \bmru{v}_1 \nonumber \\
%&\quad \left. + \left[ (\bmru{v}_2 \cdot \bmru{v}_1) (\bmru{v}_1 \cdot \bmru{v}_3)   
%  - \norm{\bmru{v}_1}^2(\bmru{v}_2\cdot\bmru{v}_3) \right] \bmru{v}_3 \right) \, . \nonumber
%  \end{align}
The vector formed by the bracket \eqref{eq:bracket} has the properties we seek with regard to the quantity $Q_i$, and appears similar to the construction for dissipating the Hamiltonian in \cite{Shashikanth10} defined using Hodge stars and musical notation.  
We may thus construct a complete set of dissipations
\begin{equation}\label{eq:dissipation}
	\bmru{d} = - \sum_{i=1}^m \epsilon_i [[Q_1, \ldots , Q_m]]_i \, ,
%	\bmru{d} = - \sum_{i=1}^m \epsilon_i [[\bmru{v}_1, \ldots , \bmru{v}_m]]_i \, ,
\end{equation}
with $\epsilon_i$ some constant coefficients. 
A little algebra shows that 
\begin{equation}
	\dot{Q}_i = \bmru{v}_i \cdot \bmru{d}= - \epsilon_i V_m^2/m! \, , 
\end{equation} 
where $V_m$ is the hypervolume of the $m$-parallelotope generated by the $\bmru{v}_i$.
We note in passing that %$\bmru{v}^i = (m!/V_m^2)[[\bmru{v}_1,\ldots,\bmru{v}_m]]_i$
$\bmru{v}^i = (m!/V_m^2)[[Q_1,\ldots,Q_m]]_i$ is the reciprocal vector of $\bmru{v}_i$, such that $\bmru{v}^j \cdot \bmru{v}_i = \delta^j_i$. %$V_m/m!$ is that of the $m$-simplex spanned 
Note also that $V_m^2 = \mathrm{det}(\{\bmru{G}\})$, where $\{\bmru{G}\}$ is the Gram matrix \cite{Shilov1977Linear} whose $ij$th entry is given by $\bmru{v}_i \cdot \bmru{v}_j$.
Each element in the sum \eqref{eq:dissipation} acts only on a single quantity, and clearly the corresponding $\epsilon_i$ coefficient can be chosen to make the quantity $Q_i$ increase or decrease, or change its magnitude if it is positive or negative semidefinite.
More importantly, the proportionality of the rates gives rise to $m-1$ independent equalities of the form
\begin{equation}
	\dot{Q}_i/\epsilon_i = \dot{Q}_j/\epsilon_j \, ,
\end{equation}
for any $i$ and $j$, and thus $m-1$ independent conserved quantities as linear combinations of pairs of the original $m$, generalizing the result in \cite{Hannaintegrable} involving a single quantity.
A convenient arbitrary choice of these constants is $R_i \equiv \left( \epsilon_{i+1}Q_i -\epsilon_{i}Q_{i+1} \right) / \sqrt{\epsilon_i^2 + \epsilon_{i+1}^2}\,$. 
%that avoids some singularities and retains the magnitude is 
The coefficients set the ratios of the rates, and any of the original $Q_i$ can be preserved by setting its associated $\epsilon_i$ to zero.  
%\begin{equation}\label{eq:define_ri}
%  R_i = \frac{\epsilon_{i+1}Q_i -\epsilon_{i}Q_{i+1}}{\sqrt{\epsilon_i^2 + \epsilon_{i+1}^2}}
%\end{equation}
The construction can be iterated by, for example, treating these $R_i$ as initial conserved quantities, %obtaining new gradients as linear combinations of the old, 
%$d_{\bmru{x}} R_i = (\epsilon_{i+1} d_{\bmru{x}} Q_i - \epsilon_i d_{\bmru{x}} Q_{i+1}) / \sqrt{\epsilon_{i}^2+\epsilon_{i+1}^2}$
%and so on, 
or by simply choosing some subset of combinations of several gradients to achieve different dynamics of any order of integrability.
It is also possible to construct modified dynamics without the volume factor $V_m^2$ by defining $\bmru{d}$ as a linear combination of the reciprocals $\bmru{v}^i$, so that the $Q_i$ change at constant rates.  However, this leads to singular behavior of $\bmru{d}$ as the volume vanishes.

An easier numerical implementation that sidesteps the need for permutations, which rapidly proliferate at large $m$, is to compute $\bmru{d}$ by the following steps. Form a column vector $\bm{\epsilon}$ from the $\epsilon_i$ coefficients, and arrange the $\bmru{v}_i$ as columns of a matrix $\{\bmru{V}\} = \{\bmru{v}_1, \bmru{v}_2, \ldots, \bmru{v}_m\}$, such that the Gram matrix is $\{\bmru{G}\} = \{\bmru{V}\}^\mathrm{T}\{\bmru{V}\}$. By considering the derivative of the determinant $\mathrm{det}(\{\bmru{G}\})$, one finds that the column vector $\bmru{d} = -\frac{1}{m!} \det(\{\bmru{V}\}^\mathrm{T}\{\bmru{V}\}) \, \{\bmru{V}\} (\{\bmru{V}\}^\mathrm{T}\{\bmru{V}\})^{-1} \bm{\epsilon}\,$. The matrix $\{\bmru{V}\} ( \{\bmru{V}\}^\mathrm{T}\{\bmru{V}\} )^{-1}$ may be recognized as the right pseudoinverse of $\{\bmru{V}\}^{\mathrm{T}}$.  To avoid singularities in this expression, it may be replaced by the tamer 
\begin{equation}\label{eq:tamer}
	\bmru{d}= -\frac{1}{m!} \, \{\bmru{V}\}\, \mathrm{adj}(\{\bmru{V}\}^\mathrm{T}\{\bmru{V}\}) \, \bmru{\epsilon} \, , 
\end{equation}
where $\mathrm{adj}$ indicates the adjugate (the transpose of the cofactor matrix).

The dissipation $\bmru{d}$ may vanish, restoring the original dynamics and conserved quantities, for either of two reasons.  One is that the system has reached an extremum of any of the modified $Q_i$, so that one of the $\bmru{v}_i$ vanishes.  The other is that two or more of the $\bmru{v}_i$ become linearly dependent, an outcome observed in the following examples. \\
%Furthermore, since the dissipation rates are proportional to the hypervolume $V_m$, we conclude that $\dot{Q}_i =0$ when at least two of the vectors $\bmru{v}_i$ become aligned (linearly dependent). 

\noindent \textit{An example where $n=3$, $m=2$} was considered in detail in \cite{Hannaintegrable}, namely the Euler equations for rigid body rotation in terms of the three components of angular momentum in a body-fixed frame, $\dot{\bm{L}} = \bm{L} \times \bm{\omega} + \bmru{d}$, where angular momentum $\bm{L}$ and angular velocity $\bm{\omega}$ are linearly related by a constant tensor, and $\bmru{d}$ may be constructed
in terms of double cross products of these (pseudo)vectors. 
In the present language, $Q_1 = \bm{L}\cdot\bm{\omega}$ is twice the energy, $Q_2 = \bm{L}\cdot\bm{L}$ is the squared angular momentum, $\bmru{v}_1 = 2\bm{\omega}$, and $\bmru{v}_2 = 2\bm{L}$. %and the dissipation coefficients in \cite{Hannaintegrable} are identified as $4\epsilon_1 = \epsilon_H$, and $4\epsilon_2 = \epsilon_L$.  
The damped system is driven to align $\bm{L}$ and $\bm{\omega}$ at a fixed point, and a conserved quantity proportional to $\epsilon_1 Q_2 - \epsilon_2 Q_1$ defines a family of quadric surfaces on which the new dynamics is confined.  %allows the attractor to be identified from initial conditions.
Although neither gradient vanishes, any of these nontrivial states has something of an extremal character, as it either maximizes energy for a given squared momentum or \emph{vice versa}, although neither quantity is fixed in the dynamics.\\

\noindent \textit{An example where $n=6$, $m=3$} is the cyclic three-particle one-dimensional Toda lattice, an integrable nonlinear system admitting solitons.
Employing Flaschka's coordinates \cite{Flaschka1974Toda}, the six undamped equations of motion are
\begin{equation}\label{eq:toda-flaschka}
\begin{aligned}
  \dot{a}_i &= a_i(b_{i+1}-b_i) \, ,\\
  \dot{b}_i &= 2(a_{i}^2-a_{i-1}^2) \, .  %; \qquad i=1,\ldots,3
\end{aligned}
\end{equation}
%$i \in {1, 2, 3}$ 
%with $i$ defined modulo 3. 
%where it is understood that $b_4 = b_1$ and $a_0=a_3$. 
Three independent first integrals are obtained \cite{Flaschka1974Toda} from invariants of the Lax matrix 
\begin{equation}\label{eq:matrixL}
 \{\ell\} = \begin{Bmatrix}
  b_1 & a_1 & a_3 \\
  a_1 & b_2 & a_2 \\
  a_3 & a_2 & b_3 
  \end{Bmatrix} \, ,
\end{equation}
namely $Q_1 = \mathrm{tr}(\{\ell\})$, $Q_2 = \mathrm{tr}(\{\ell\}^2)$, and $Q_3 = \mathrm{tr}(\{\ell\}^3)$, where $\mathrm{tr}$ is the trace operation. 
 The quantity $Q_2$ is proportional to the energy.
 This system is actually superintegrable, admitting additional constants of the motion beyond the three considered here \cite{Agrotis06}, but presumably our modified dynamics will violate these while creating two integrals from the explicitly considered original three. 
 %Letting $\bmru{x} = (a_1, a_2,a_3, b_1, b_2, b_3)$, the gradients $\bmru{v}_i = d_{\bmru{x}}Q_i$ are $\bmru{v}_1 = (0,0,0,1,1,1)$, $\bmru{v}_2 = 2(2a_1,2a_2,2a_3,b_1,b_2,b_3)$, and $\bmru{v}_3 =  3(2 a_2 a_3 + 2 a_1 (b_1 + b_2), 2 a_3 a_1 + 2 a_2 (b_2 + b_3), 2 a_1 a_2 + 2 a_3 (b_3 + b_1), a_3^2 + a_1^2 + b_1^2,  a_1^2 + a_2^2 + b_2^2,  a_2^2 + a_3^2 + b_3^2 )$.
  We note that, unlike the original system \eqref{eq:toda-flaschka}, our modified dynamics allow for unphysical zero-crossings of the $a_i$, so in general some care may be required either in choosing variables or defining the dissipations.
%\begin{equation}\label{eq:vectors_ui}
%\begin{aligned}
%  \bmru{v}_1 = \partial_{\bmru{x}} Q_1 = & [0,0,0,1,1,1]^{\mathrm{T}};\\
%  \bmru{v}_2 = \partial_{\bmru{x}} Q_2 = & 2[2a_1,2a_2,2a_3,b_1,b_2,b_3]^{\mathrm{T}};\\
%  \bmru{v}_3 = \partial_{\bmru{x}} Q_3 = &
%  3[2 a_2 a_3 + 2 a_1 (b_1 + b_2), 2 a_3 a_1 + 2 a_2 (b_2 + b_3), 2 a_1 a_2 +
% 2 a_3 (b_3 + b_1), \ldots\\ & (a_3^2 + a_1^2 + b_1^2),  (a_1^2 + a_2^2 +
%   b_2^2),  (a_2^2 + a_3^2 + b_3^2)]^{\mathrm{T}};
%\end{aligned}
%\end{equation}

\begin{figure}[h!]
	\includegraphics[width=3in]{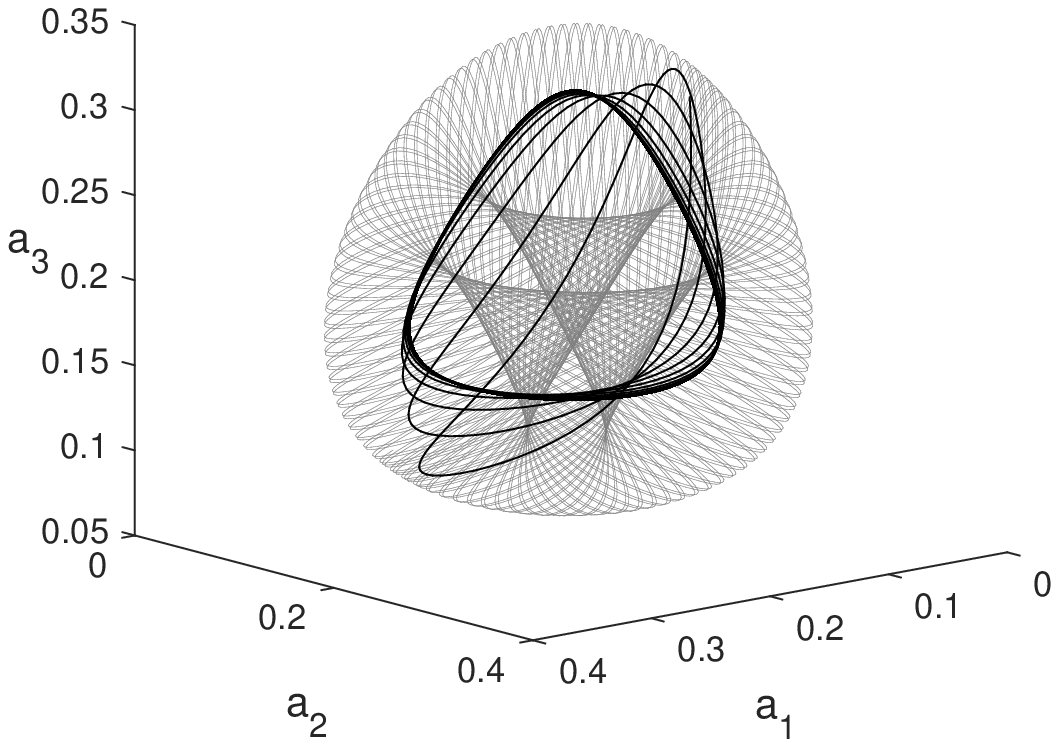}
	\includegraphics[width=3in]{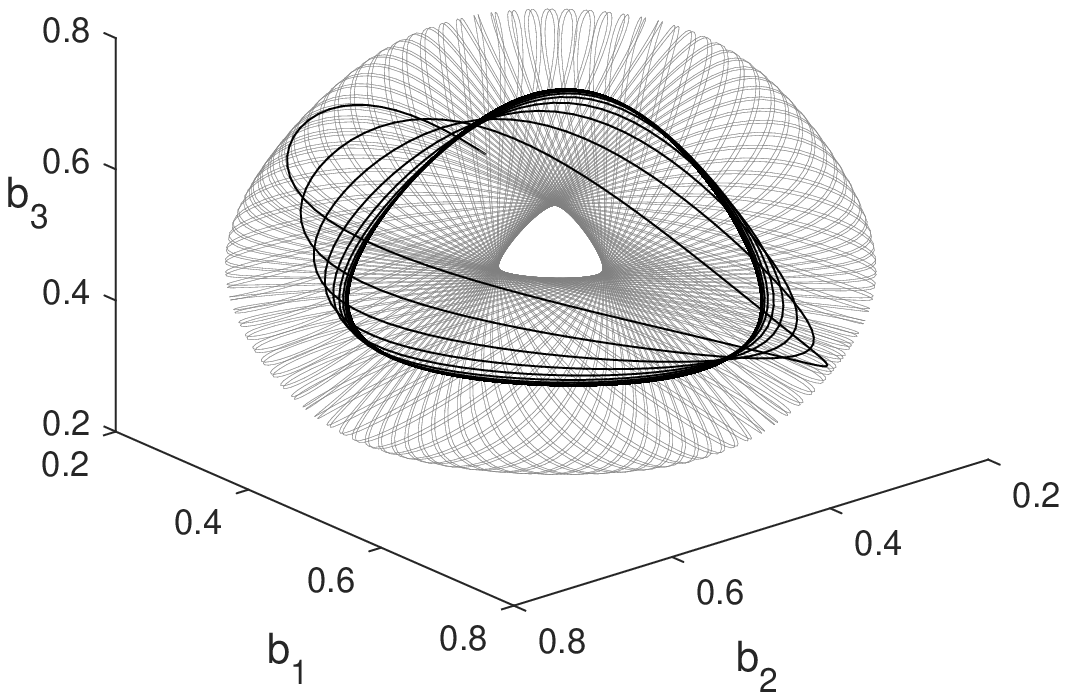}
	\caption{Trajectories of the six Flaschka variables for the cyclic three-body Toda lattice \eqref{eq:toda-flaschka} (grey) and the modified system with  $( \epsilon_1, \epsilon_2, \epsilon_3 ) = (0.075, 0.05, 0.025)$ (black) up to 1000 time units.  Initial conditions are $(a_1, a_2, a_3, b_1, b_2, b_3) = (0.1, 0.2, 0.3, 0.4, 0.5, 0.6)$.}\label{todatrajectories}
\end{figure}

The evolution of the original and modified system is shown in Figure \ref{todatrajectories} for a particular choice of initial conditions and dissipations. 
Without damping, the trajectories appear low-dimensional and weakly aperiodic, possibly quasiperiodic. 
 Addition of damping drives the system towards a limit cycle in which the three original quantities take on new steady values, and the three gradients are linearly dependent.   
%limit cycle may be a soliton of original system.  don't know.  did we drive the system from two solitons to one?where is the radiation?
Note that $Q_1$ defines a %n ``octahedral'' 
   plane in $b$-space; both the original trajectory and the limit cycle lie on different parallel planes in this subspace.
In this example, $\bmru{v}_1$ is actually a constant and so cannot shrink or realign, but 
   %  \bmru{v}_1 = \partial_{\bmru{x}} Q_1 = & [0,0,0,1,1,1]^{\mathrm{T}};\\
$\bmru{v}_2$ and $\bmru{v}_3$ evolve such that the volume of the parallelepiped generated by the $\bmru{v}_i$ shrinks rapidly to zero.  While the changes in the $Q_i$ are monotonic by construction, that of the volume is not.

\begin{figure}[h!]
	\includegraphics[width=6in]{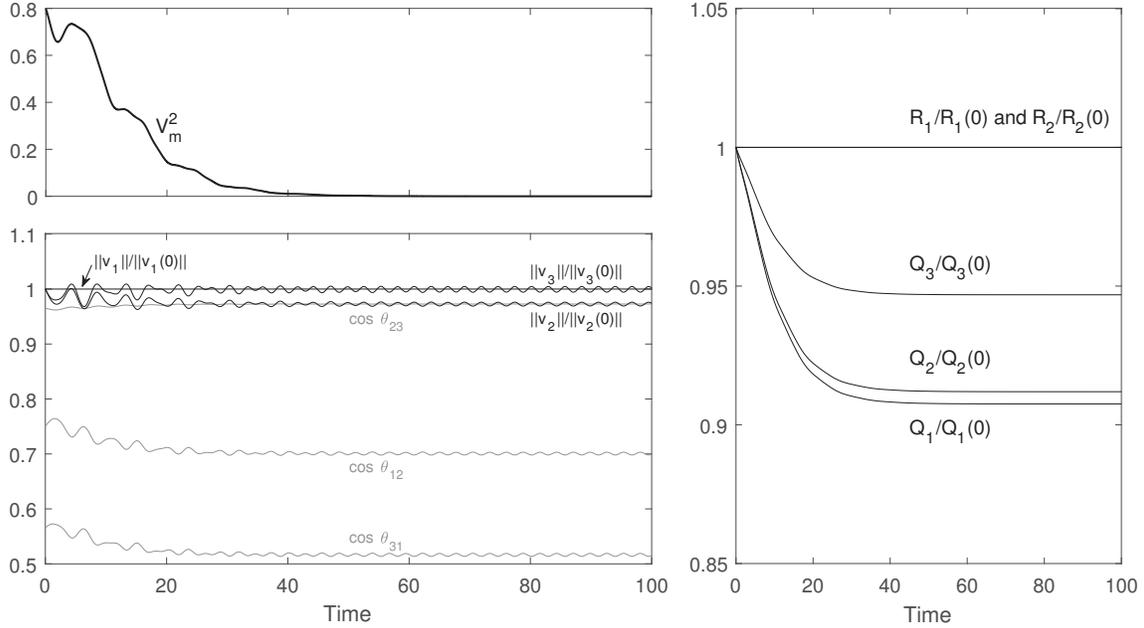}
	\caption{Evolutions during the initial phase of the modified trajectory from Figure \ref{todatrajectories}.  Shown are the squared volume $V^2_m$ of the parallelepiped generated by the three gradients $\bmru{v}_i$, their magnitudes $\norm{\bmru{v}_i}$ normalized by initial values (black), the cosines of their included angles $\theta_{ij}$ (grey), and the monotonically decreasing formerly conserved quantities $Q_i$ and presently conserved quantities $R_i$ normalized by initial values.  The volume and associated dissipations rapidly vanish as the three gradients become coplanar, with no appreciable shrinking or pairwise alignment of gradients.}\label{todadata}
\end{figure}

The extension to three conserved quantities illustrates an interesting point, in that it is not pair- or group-wise alignment that is induced by the dissipation, but linear dependence, here in the form of coplanarity.  This can be gleaned from the various evolutions shown in Figure \ref{todadata}. 
The magnitudes of the $\bmru{v}_i$ change very little, and there is no pairwise alignment, as seen from the cosines $\cos\theta_{ij} \equiv \bmru{v}_i \cdot \bmru{v}_j / ( \norm{\bmru{v}_i} \norm{\bmru{v}_j} )$, none of which are driven to unity.  Thus, the rapid change in volume $V_m$ comes from a collective linear dependence of all three vectors.\\
% in the form of the vanishing of the quantity $1+2\cos\theta_{12}\cos\theta_{23}\cos\theta_{31}-\cos^2\theta_{12}-\cos^2\theta_{23}-\cos^2\theta_{31}$, the part of the expression for $V^2_m$ that is independent of the magnitudes of the $\bmru{v}_i$.\\
%Integrated with ode45, relative tol = 1e-8, absolute tol = 1e-10

\newpage In general, while a system can return to its unmodified form by approaching an extremum characterized by a vanishing gradient, the restoring outcomes we often tend to observe are instead the result of the degeneration of the $m$-parallelotope into a linearly dependent set of gradients spanning an $(m-1)$-dimensional subspace.
At this point one can construct a new set of $m-2$ gradients from the old, and start a new process.  When only one gradient remains, the dissipation process affects all quantities and can only halt by reaching an extremal point where this gradient vanishes.
We reiterate that the modified dynamics need not be constrained to decay terms, but can include amplification of any number of the original quantities.

When $m=2$, the linear dependence simply corresponds to the alignment of two vectors, as in damped rigid body rotation \cite{Hannaintegrable} or the Beltramization of fluid flows under selective energy dissipation \cite{Vallis89}.
In the rigid body, alignment also corresponds to extrema of one quantity on the particular final level sets reached.
It is an interesting question whether this latter property has any generalization to other systems, particularly those with $m > 2$.  %Of course iterating a dissipative process with successively fewer gradients will eventually lead to a system that will only come to a stop when a single remaining $\bmru{v}$ shrinks to zero and a 
%energy exchange/transfer

The bracket \eqref{eq:bracket} can be rearranged and interpreted as a linear operator acting on the subscripted gradient, to facilitate comparison with other forms inspired by thermodynamics \cite{Morrison86, GrmelaOttinger97},  
%((bi)linearity already mentioned in Morrison84)
with \eqref{eq:tamer} perhaps providing a more convenient access point than \eqref{eq:dissipation}.
For example, the particular bracket written explicitly earlier is of the form 
%$[[\bmru{v}_1,\bmru{v}_2,\bmru{v}_3]]_2 $
$[[Q_1,Q_2,Q_3]]_2 = \frac{1}{6} \Big( \Big[ \norm{\bmru{v}_1}^2 \norm{\bmru{v}_3}^2 - (\bmru{v}_3 \cdot \bmru{v}_1)^2 \Big] \bmru{I} \\
- \Big[ \norm{\bmru{v}_3}^2 \bmru{v}_1 - (\bmru{v}_1 \cdot \bmru{v}_3) \bmru{v}_3 \Big] \bmru{v}_1
- \Big[ \norm{\bmru{v}_1}^2 \bmru{v}_3 - (\bmru{v}_3 \cdot \bmru{v}_1) \bmru{v}_1 \Big] \bmru{v}_3 \Big) \cdot \bmru{v}_2$, where $\bmru{I}$ is the $n$-dimensional identity. 
It is illustrative to compare this with a corresponding term that simply projects the gradient so as to remove any components that would act on other quantities, a process analogous to the construction of the bilinear metriplectic bracket \cite{Morrison86}, 
$ \Big( \bmru{I} - \norm{\bmru{v}_1}^{-2}\bmru{v}_1\bmru{v}_1  -  \norm{\bmru{v}_3}^{-2}\bmru{v}_3\bmru{v}_3  \Big) \cdot \bmru{v}_2$.  
The difference in weights is important, as a combination of projective terms will not conserve any quantities, and will drive the system towards a trivial equilibrium (as may be easily confirmed by its application to a system such as the Toda example above), while the exterior dissipative mechanism conserves a number of quantities less by one than the original system. 

As a tool of discovery, this process of modifying dynamics should be useful in the construction of new integrable systems or the search for nontrivial states of known systems.
Given further information about a particular system to be modified, the process is also potentially a tool of control with which to set not only the rates, but the overall values of the original quantities, or steer the trajectory of the system towards particular states.
 Future avenues that immediately suggest themselves include extensions to curved spaces, integrable PDEs, and discrete-time dynamics.

\section*{ACKNOWLEDGMENTS}

We thank V. Fedonyuk and A. G. Nair for conversations.

\bibliographystyle{unsrt}
%\bibliography{refs_dissipation}

\end{document}